\documentclass{article}
\usepackage{spconf,amsmath,graphicx}

\newcommand{\bi}[1]{\ensuremath{\boldsymbol{#1}}}   %  Definition of Bold Itaric
\usepackage{amssymb}
\usepackage{enumerate}
\usepackage{graphicx}
\usepackage{multirow}
\usepackage{arydshln}
\usepackage{color}
\usepackage{float}
\usepackage{bm}
\usepackage{comment}
\usepackage{cite}
\usepackage{url}
\usepackage{amsmath}
\usepackage{algorithmic}
\usepackage{array}
\usepackage{colortbl}
\usepackage{subfigure}

\newlength\savedwidth
\newcommand{\wcline}[1]{\noalign{\global\savedwidth\arrayrulewidth\global\arrayrulewidth 1.0pt} \cline{#1}
\noalign{\global\arrayrulewidth\savedwidth}}

\title{Sound Event Detection Utilizing Graph Laplacian Regularization\\with Event Co-occurrence}
\name{Keisuke Imoto$^\dagger$ and Seisuke Kyochi$^\ddagger$}
\address{$^\dagger$Ritsumeikan University, Japan, $^\ddagger$University of Kitakyusyu}
\begin{document}
%\ninept
%
\maketitle
%
%---------------------------------------------------
\begin{abstract}
%---------------------------------------------------
A limited number of types of sound event occur in an acoustic scene and some sound events tend to co-occur in the scene; for example, the sound events ``dishes'' and ``glass jingling'' are likely to co-occur in the acoustic scene ``cooking.''
In this paper, we propose a method of sound event detection using graph Laplacian regularization with sound event co-occurrence taken into account.
In the proposed method, the occurrences of sound events are expressed as a graph whose nodes indicate the frequencies of event occurrence and whose edges indicate the sound event co-occurrences.
This graph representation is then utilized for the model training of sound event detection, which is optimized under an objective function with a regularization term considering the graph structure of sound event occurrence and co-occurrence.
Evaluation experiments using the TUT Sound Events 2016 and 2017 detasets, and the TUT Acoustic Scenes 2016 dataset show that the proposed method improves the performance of sound event detection by 7.9 percentage points compared with the conventional CNN-BiGRU-based detection method in terms of the segment-based F1 score.
In particular, the experimental results indicate that the proposed method enables the detection of co-occurring sound events more accurately than the conventional method.
%
%---------------------------------------------------
\end{abstract}
%---------------------------------------------------
%
%---------------------------------------------------
%---------------------------------------------------
%------------------------------------------------------------------
%\vspace{0pt}
\section{Introduction}
\label{sec:intro}
%\vspace{0pt}
%------------------------------------------------------------------
Sound event detection (SED) is a task that identifies types of sound and detects their onset and offset \cite{Imoto_AST2018_01}.
Recently, many works have addressed SED because SED has a large potential for many applications such as monitoring elderly people or infants \cite{Peng_ICME2009_01,Guyot_ICASSP2013_01}, automatic surveillance \cite{Radhakrishnan_WASPAA2005_01,Ntalampiras_ICASSP2009_01,Chandrakala_CSUR2019_01}, automatic anomaly detection \cite{Koizumi_TASLP2019_01,Kawaguchi_ICASSP2019_01}, and media retrieval \cite{Jin_INTERSPEECH2012_01}.

SED is typically categorized into two types: monophonic and polyphonic SED.
In monophonic SED, it is assumed that multiple sound events do not occur simultaneously; thus, a monophonic SED system only detects at most one sound event at a time.
However, in a real environment, since multiple sound events often occur simultaneously, monophonic SED shows limited performance in a real-life situation.
To address this problem, polyphonic SED systems, which can detect multiple overlapping sound events in time, have been developed.

One approach to polyphonic SED is non-negative matrix factorization (NMF) \cite{Dessein_MIG2013_01,Komatsu_DCASE2016_01}.
In the NMF-based SED approach, a sound with polyphonic events is decomposed into a product of a basis and activation matrices, where each basis vector and activation vector respectively represent a single sound event and the active duration of the corresponding sound event.
More recently, neural network-based SED approaches have also been widely developed.
For example, a convolutional neural network (CNN)-based approach, which can detect sound events robustly against time and frequency shifts in the input acoustic feature, has been used in many works \cite{Hershey_ICASSP2017_01,Jeong_DCASE2017_01}.
Recurrent neural network (RNN)- or convolutional recurrent neural network (CRNN)-based approaches, which can capture temporal information of sound events, have also been utilized in some works \cite{Cakir_TASLP2017_01,Hayashi_TASLP2017_01,Kothinti_DCASE2018_01}.
These methods successfully analyze overlapping sound events with reasonable performance.
However, when the number of types of sound events increases, these neural network-based approaches require a large training dataset.

\begin{figure*}[t]
\centering
\vspace{-2pt}
\includegraphics[width=1.96\columnwidth]{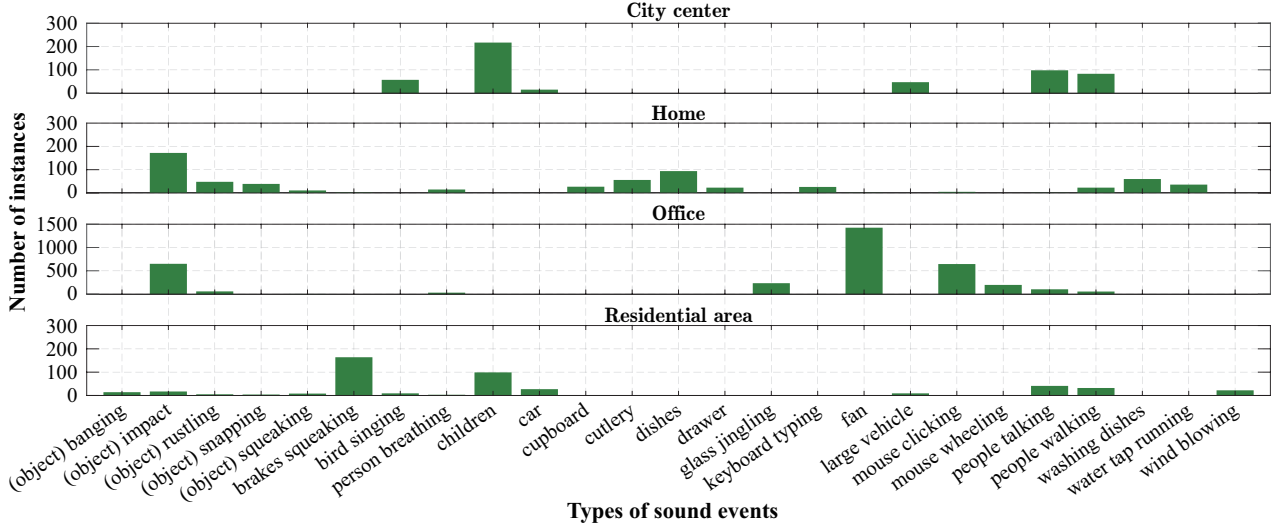}
\vspace{-6pt}
\caption{Frequency of sound event instances for each acoustic scene in dataset used for evaluation experiments}
\label{fig:numofinstance}
\end{figure*}

The number of types of sound events occurring in a single acoustic scene is limited and some sound events tend to co-occur, as shown in Fig.~\ref{fig:numofinstance}.
For example, the sound events ``dishes'' and ``cutlery'' are likely to co-occur, and ``car'' and ``brakes squeaking'' also tend to co-occur.
By considering this co-occurrence in the model training of sound events, we expect to be able to model sound events efficiently and effectively with a limited amount of sound data.
On the basis of this idea, Mesaros \textit{et al.} \cite{Mesaros_EUSIPCO2011_01} and Imoto and Ono \cite{Imoto_TASLP2019_01} have respectively proposed methods of SED and acoustic scene classification with the co-occurrence of sound events taken into account, which were based on Bayesian generative models.
However, conventional methods cannot be integrated into the state-of-the-art neural network-based methods.

%
%\vspace{0pt}

To address this limitation, we have proposed a neural network-based method for SED that can consider the co-occurrence of sound events in each sound clip \cite{Imoto_ICASSP2019_01}.
To take the co-occurrence of sound events into account, we introduce graph Laplacian regularization with event co-occurrence into the objective function of a neural network.
The proposed method with graph Laplacian regularization is implemented using a graph representation of a sound event occurrence and the event co-occurrence.
The graph representation of the sound event co-occurrence can be constructed using prior information on the event co-occurrence or only from the dataset used for the model training of the neural network.
Thus, the proposed method does not require additional training data that are different from the data used in conventional machine learning-based methods \cite{Hershey_ICASSP2017_01,Jeong_DCASE2017_01,Cakir_TASLP2017_01,Hayashi_TASLP2017_01,Kothinti_DCASE2018_01,Mesaros_EUSIPCO2011_01}.
In this paper, we will discuss in detail the proposed method using graph Laplacian regularization with event co-occurrence and perform a detailed evaluation of the behavior of SED using graph Laplacian regularization.

The remainder of this paper is structured as follows.
In section 2, we introduce the conventional SED approach based on a CRNN.
In section 3, we discuss the proposed approach to SED, in which the co-occurrence of sound events can be considered.
In section 4, we report experiments conducted to evaluate the performance of SED by the proposed and conventional methods,
and in section 5, we  conclude this paper.
%
%
%
%------------------------------------------------------------------
\vspace{-10pt}
\section{Conventional Sound Event Detection Based on Convolutional Recurrent Neural Networks}
\label{sec:crnn}
\vspace{0pt}
%------------------------------------------------------------------
In this section, we present an overview of neural network-based SED approaches.
For polyphonic SED, many researches apply CNN architectures \cite{Hershey_ICASSP2017_01,Jeong_DCASE2017_01}.
In CNN-based SED, the time-frequency representation of an acoustic signal ${\bf V} \in \mathbb{R}^{D \times T}$ is input to a convolutional layer, where $D$ and $T$ are the dimension of the acoustic feature and the number of time frames of the acoustic feature, respectively.
The CNN layer convolutes the acoustic feature map with two-dimensional filters; after that, max pooling is operated to reduce the dimension of the feature map.
The CNN architecture allows feature extraction robust against time and frequency shifts, which often occur in environmental sound analysis.

An RNN has also been applied to SED in some works \cite{Cakir_TASLP2017_01,Hayashi_TASLP2017_01,Kothinti_DCASE2018_01} to explicitly model time correlations of sound events.
In particular, it has been reported that neural networks combining the CNN and a bidirectional gated recurrent unit (BiGRU) \cite{Schuster_TSP1997_01,Cho_arXiv2014_01}, which can capture forward and backward temporal correlations of sound events, successfully detected sound events.
In the SED based on CNN-BiGRU, the acoustic feature map ${\bf V}$ is fed to the convolutional layer.
The output of the convolutional layer in the $t$th time frame ${\bf x}_{t}^{(d, c)}$ is then concatenated as 
${\bf x}_{t} = (x_{t}^{(1,1)}\!\!, \ x_{t}^{(1,2)}\!\!, \ \ldots, x_{t}^{(1,C)}\!\!, \ x_{t}^{(2,1)}\!\!, \ \ldots, x_{t}^{(d,c)}\!\!, \ \ldots, x_{t}^{(D',C)})$, where $C$ is the number of filters of the convolution layer.
After that, ${\bf x}_t$ is fed to the BiGRU layer, and the output vector ${\bf h}_{t}$ of the BiGRU layer is calculated using the following equations:

\vspace{-8pt}
\begin{align}
{\bf g}^{f}_{t} &= \sigma({\bf W}^{f}_{g} {\bf x}_{t} + {\bf U}^{f}_{g} {\bf h}_{t-1} + {\bf b}^{f}_{g}),\\[1pt]
{\bf r}^{f}_{t} &= \sigma({\bf W}^{f}_{r} {\bf x}_{t} + {\bf U}^{f}_{r} {\bf h}_{t-1} + {\bf b}^{f}_{r}),\\[1pt]
{\bf h}^{f}_{t} &= (1-{\bf g}^{f}_{t}) \odot {\bf h}_{t-1} \nonumber\\[0pt]
&\hspace{10pt}+ {\bf g}^{f}_{t} \odot \tanh ({\bf W}^{f}_{h} {\bf x}_{t} + {\bf U}^{f}_{h} ( {\bf r}^{f}_{h} {\bf h}_{t-1}) + {\bf b}^{f}_{h}),\\[2pt]
{\bf g}^{b}_{t} &= \sigma({\bf W}^{b}_{g} {\bf x}_{t} + {\bf U}^{b}_{g} {\bf h}_{t+1} + {\bf b}^{b}_{g}),\\[1pt]
{\bf r}^{b}_{t} &= \sigma({\bf W}^{b}_{r} {\bf x}_{t} + {\bf U}^{b}_{r} {\bf h}_{t+1} + {\bf b}^{b}_{r}),\\[1pt]
{\bf h}^{b}_{t} &= (1-{\bf g}^{b}_{t}) \odot {\bf h}_{t+1} \nonumber\\[0pt]
&\hspace{10pt}+ {\bf g}^{b}_{t} \odot \tanh ({\bf W}^{b}_{h} {\bf x}_{t} + {\bf U}^{b}_{h} ( {\bf r}^{b}_{h} {\bf h}_{t+1}) + {\bf b}^{b}_{h}),\\[1pt]
{\bf h}_{t} &=
\begin{bmatrix}
{\bf h}^{f}_{t}\\[1pt]
{\bf h}^{b}_{t}
\end{bmatrix},
\label{eq:bigru}
\end{align}
%\vspace{0pt}
%

\noindent where superscripts $f$ and $b$ indicate the forward and backward networks, respectively.
Subscripts $t$, $g$, and $r$ indicate the time index, update gate, and reset gate, respectively.
${\bf g}$, ${\bf r}$, $\odot$, and $\sigma$ indicate the update gate vector, reset gate vector, Hadamard product, and sigmoid function, respectively.
${\bf W}$ and ${\bf U}$ are parameter matrices and ${\bf b}$ is a bias vector.
The BiGRU layer is followed by a fully connected layer, which is the output layer of the network.
The final output of the network is calculated as

%
%\vspace{0pt}
\begin{align}
{\bf y}_{t} &= \sigma({\bf h}_{t}).
\label{eq:outputlayer}
\end{align}
%\vspace{0pt}
%

The parameters of the CNN-BiGRU network for SED are optimized under the following sigmoid cross-entropy objective function $E({\bi \Theta})$ using the backpropagation through time (BPTT) \cite{Werbos_IEEE1990_01}:

%
%\vspace{0pt}
\begin{align}
\hspace{-0pt} E({\bi \Theta}) &= - \sum^{T}_{t=1} {\big \{} {\bf z}_{t} \log ( {\bf y}_{t} ) + (1-{\bf z}_{t}) \log (1-{\bf y}_{t}) {\big \}}\nonumber \\
&= - \hspace{-2.5pt} \sum^{M}_{m=1} \hspace{-1pt} \sum^{T}_{t=1} \hspace{-1pt} {\Big \{} \hspace{-1.5pt} z_{m,t} \log y_{m,t} + \hspace{-1pt} (1 \hspace{-1.25pt} - \hspace{-1.25pt} z_{m,t}) \log \hspace{-1pt} {\big (} 1 \hspace{-1.25pt} - \hspace{-1.25pt} y_{m,t} {\big )} \hspace{-1.25pt} {\Big \}},
\label{eq:crnn_objective}
%\vspace{0pt}
\end{align}

\noindent where ${\bf z}_{m,t}$ is a target vector of the output that indicates whether sound events are active or nonactive in time frame $t$.
$M$ indicates the number of types of sound events.
%
%
%
%\vspace{0pt}
\begin{figure}[t]
\vspace{5pt}
\centering
%\vspace{0pt}
\includegraphics[width=0.999\columnwidth]{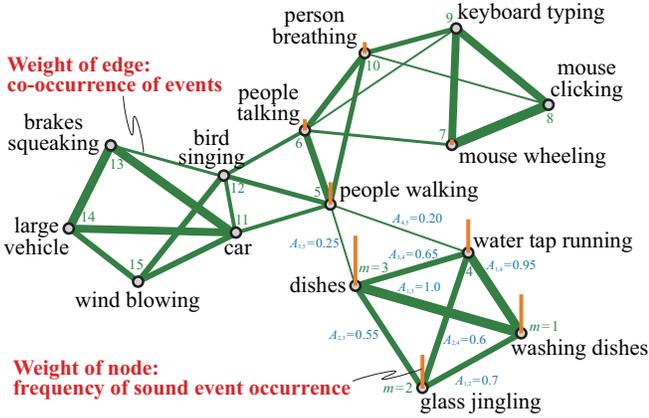}
%\vspace{0pt}
\caption{Example of graph representation of sound event occurrences}
\label{fig:adjacent01}
\end{figure}
%
%
%
%------------------------------------------------------------------
%\vspace{0pt}
\section{Sound Event Detection with Event-co-occurrence-based Regularization}
\label{sec:proposed}
%\vspace{0pt}
%------------------------------------------------------------------
%- - - - - - - - - - - - - - - - - - - - - - - - - - - - - - - - - - - -
%\vspace{0pt}
\subsection{Motivation}
\label{subsec:motivation}
%\vspace{0pt}
%- - - - - - - - - - - - - - - - - - - - - - - - - - - - - - - - - - - -
Conventional CRNN-based approaches achieve reasonable event detection performances when there is a sufficient amount of training sound data.
However, since recording and annotating environmental sounds are very time-consuming \cite{Imoto_AST2018_01}, in many situations, the conventional CRNN-based methods are likely to exhibit degradation in their event detection performance.
To address this limitation, we propose a new method of SED using graph Laplacian regularization based on sound event co-occurrence.

As shown in Fig.~\ref{fig:numofinstance}, the number of types of sound events occurring in a single acoustic scene is limited, and some sound events tend to co-occur.
For example, the sound events ``dishes'' and ``glass jingling'' tend to co-occur, and ``car'' and ``brakes squeaking'' are also likely to co-occur.
Considering the sound event co-occurrence in the model parameter estimation of a neural network, we expect that sound events can be efficiently and effectively modeled with a limited amount of sound data.
\begin{table}[t!]
%\vspace{0pt}
\caption{Experimental conditions}
\label{tab:Condition}
%\vspace{0pt}
\small
\centering
\renewcommand{\arraystretch}{1.0}
\begin{tabular}{lll}
\wcline{1-3}
\multicolumn{2}{l}{\!\!}&\vspace{-9pt} \\
\multicolumn{2}{l}{\!\!Acoustic feature}&Log mel-band energy\!\!\\
\multicolumn{2}{l}{\!\!\# dims. of acoustic feature}&64\!\!\\
\multicolumn{2}{l}{\!\!Frame length}&40 ms\!\!\\
\multicolumn{2}{l}{\!\!Frame shift}&20 ms\!\!\\
\multicolumn{2}{l}{\!\!Length of sound clip}&10 s\!\!\\
\multicolumn{2}{l}{\!\!Regularization weight $\alpha$}&$1.0 \times 10^{-5}$\!\!\\
\cline{1-3}\\[-7pt]
\multicolumn{2}{l}{\!\!Network structure of CNN-BiGRU}&3 conv. \& 1 BiGRU layers\!\!\\
\multicolumn{2}{l}{\!\!Filter size in CNN layers}&3 $\times$ 3\!\!\\
\multicolumn{2}{l}{\!\!Pooling in CNN layers}&3 $\times$ 1 max pooling\!\!\\
\multicolumn{2}{l}{\!\!Activation function}&ReLU\!\!\\
\multicolumn{2}{l}{\!\!\# channels of CNN layers}&128, 128, 128\!\!\\
\multicolumn{2}{l}{\!\!\# GRU units}&32\!\!\\
\multicolumn{2}{l}{\!\!\# epochs for training}&150\!\!\\
\multicolumn{2}{l}{\!\!Optimizer}&Adam \cite{Kingma_ICLR2015_01}\!\!\\
\multicolumn{2}{l}{\!\!Thresholding}&Adaptive thresholding \cite{Xu_DCASE2017_01}\!\!\\[2pt]
\wcline{1-3}
\end{tabular}
%\vspace{0pt}
\end{table}
\begin{table*}[t]
\small
\caption{Setting of adjacency matrix ${\bf A}$ used in evaluation experiments in Secs. 4.2 and 4.3}
%\vspace{0pt}
\label{tbl:event_occurrence}
\centering
\begin{tabular}{c}
\hspace{-8.5pt}
\includegraphics[width=2.115\columnwidth]{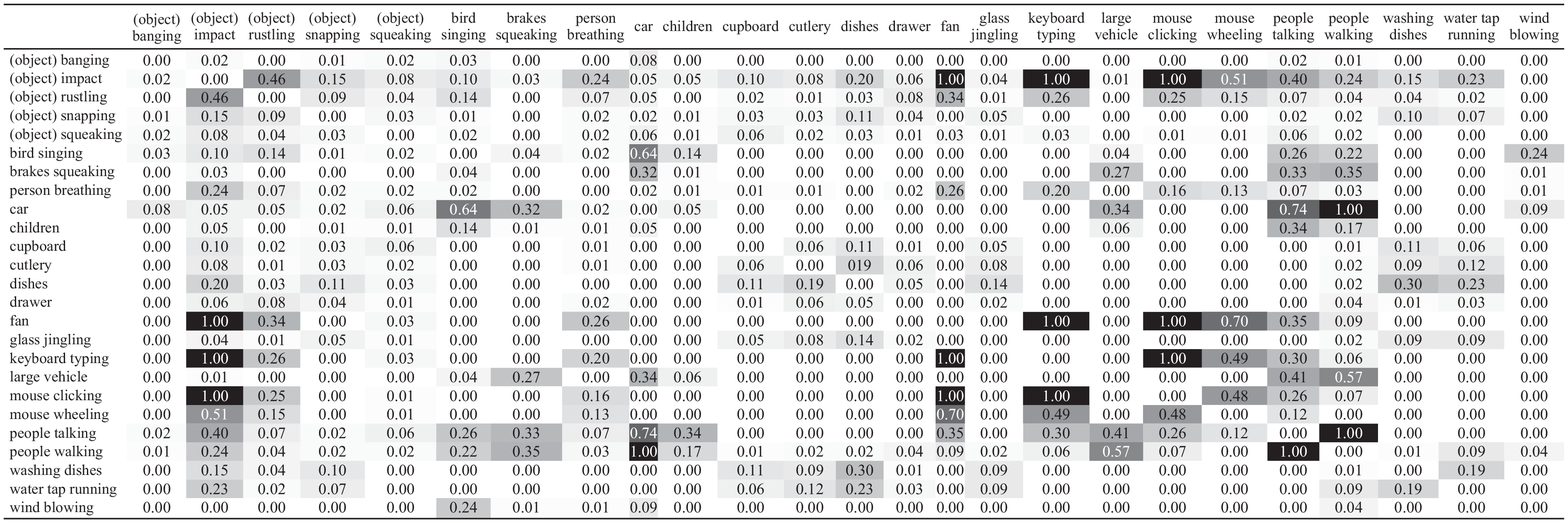}
\end{tabular}
%\vspace{0pt}
\end{table*}
\begin{table*}[t!]
%\vspace{0pt}
\caption{Detection performance for sound events in segment-based metrics}
\label{tab:Result01}
%\vspace{0pt}
\small
\centering
\renewcommand{\arraystretch}{1.0}
\begin{tabular}{lrrrrrrrrrr}
\wcline{1-11}
\!&&&&&&\vspace{-9pt} \\
\multicolumn{1}{c}{\multirow{2}{*}{\!\!\bf Method}}&\multicolumn{2}{c}{\bf Fold 1}&\multicolumn{2}{c}{\bf Fold 2}&\multicolumn{2}{c}{\bf Fold 3}&\multicolumn{2}{c}{\bf Fold 4}&\multicolumn{2}{c}{\bf Average}\!\!\!\\
\cline{2-11}\\[-9pt]
\!\!&F1 score\!\!&\!\!Error rate&F1 score\!\!&\!\!Error rate&F1 score\!\!&\!\!Error rate&F1 score\!\!&\!\!Error rate&F1 score\!\!&\!\!Error rate\!\!\!\\
\wcline{1-11}\\[-7pt]
\!\!CNN&48.67\%\!\!&\!\!0.708&31.36\%\!\!&\!\!0.829&33.11\%\!\!&\!\!0.813&23.55\%\!\!&\!\!0.899&34.17\%\!\!&\!\!0.812\!\!\!\\
\!\!CNN-GRU&51.00\%\!\!&\!\!0.672&36.64\%\!\!&\!\!0.795&35.95\%\!\!&\!\!0.797&34.70\%\!\!&\!\!0.864&39.57\%\!\!&\!\!0.782\!\!\!\\
\!\!CNN-BiGRU&53.10\%\!\!&\!\!0.652&35.10\%\!\!&\!\!0.807&38.34\%\!\!&\!\!0.769&38.42\%\!\!&\!\!{\bf 0.814}&41.24\%\!\!&\!\!0.761\!\!\!\\
\!\!CNN-BiGRU w/ GLR\!\!\!&{\bf 55.59}\%\!\!&\!\!{\bf 0.631}&{\bf 48.28\%}\!\!&\!\!{\bf 0.742}&{\bf 50.39\%}\!\!&\!\!{\bf 0.678}&{\bf 42.39\%}\!\!&\!\!0.820&{\bf 49.16}\%\!\!&\!\!{\bf 0.718}\!\!\!\\
\wcline{1-11}
\end{tabular}
%\vspace{-4pt}
\end{table*}
%
%
%
%- - - - - - - - - - - - - - - - - - - - - - - - - - - - - - - - - - - -
%\vspace{0pt}
\subsection{Sound Event Detection Using Graph Laplacian Regularization}
\label{subsec:SEDwithGLR}
%\vspace{0pt}
%- - - - - - - - - - - - - - - - - - - - - - - - - - - - - - - - - - - -
To consider the co-occurrence of sound events, we introduce a graph representation of sound event occurrences and a graph-based regularization technique for sound event modeling.

Suppose that the graph representation ${\bf G}$ of the sound event occurrence has nodes ${\bf v} \in \mathbb{R}^{M}$ and adjacency matrix ${\bf A} \in \mathbb{R}^{M \times M}$, as shown in Fig.~\ref{fig:adjacent01}.
The weights of the nodes on the graph indicate the frequencies of sound event occurrences, and the weights of the edges are how often two sound events co-occur.
The graph Laplacian matrix ${\bf L}$ \cite{Shuman_SPM2013_01} of this graph is defined as 

%
%\vspace{0pt}
\begin{align}
{\bf L} = {\bf \Delta} - {\bf A},
\end{align}
%\vspace{0pt}
%

\noindent where ${\bf \Delta}$ is a diagonal, so-called degree matrix, whose diagonal elements are defined as

%
%\vspace{0pt}
\begin{align}
\Delta_{ii} = \sum_{j} A_{i,j}.
\end{align}
%\vspace{0pt}
%
%
$A_{i,j}$ indicates the weight of the edge representing the connection between the $i$th and $j$th node.

If two sound events are likely to co-occur, that is, the edge between these sound events has a large weight, the frequencies of occurrences of the two sound events should have a small difference.
In the proposed method, we thus consider the following penalty term:

\begin{comment}
%
%\vspace{0pt}
\begin{align}
\frac{1}{2} \sum_{i, j =0}^{M} A_{i,j} \| v_{i} - v_{j} \|^{2} &= \sum_{i = 0}^{M} v_{i} v_{i} \Delta_{i,i} - \sum_{i,j = 0}^{M} v_{i} v_{j} A_{i,j}\nonumber \\[3pt]
&= \mathrm{Tr} ({\bf v}^{\mathsf T} {\bf \Delta} {\bf v}) - {\bf v}^{\mathsf T} {\bf A} {\bf v}\nonumber \\[3pt]
&= \mathrm{Tr} ({\bf v}^{\mathsf T} {\bf L} {\bf v}),
\label{eq:term1}
\end{align}
%\vspace{0pt}
%
\end{comment}
%
%\vspace{0pt}
\begin{align}
\frac{1}{2} \sum_{i, j =0}^{M} A_{i,j} \hspace{1pt} (v_{i} - v_{j} )^{2} &= \sum_{i = 0}^{M} v_{i} v_{i} \Delta_{i,i} - \sum_{i,j = 0}^{M} v_{i} v_{j} A_{i,j}\nonumber \\[3pt]
&= {\bf v}^{\mathsf T} {\bf \Delta} {\bf v} - {\bf v}^{\mathsf T} {\bf A} {\bf v}\nonumber \\[3pt]
&= {\bf v}^{\mathsf T} {\bf L} {\bf v},
\label{eq:term1}
\end{align}
%\vspace{0pt}
%

\noindent where $v_{i}$ is the weight of node $i$.
This penalty term induces the co-occurrence of sound events which has the edge with a large weight, by adding a large penalty if the two sound events have a large difference.
Adding this penalty to the objective function in the neural network for SED enables us to learn a sound event model in which we can consider the sound event co-occurrence \cite{Cai_TPAMI2011_01,Ichita_APSIPA2018_01}.
\noindent By substituting Eq. (\ref{eq:term1}) into Eq. (\ref{eq:crnn_objective}), we obtain the following objective function:

%
%\vspace{0pt}
\begin{align}
E({\bi \Theta}) &= - \sum^{T}_{t=1} {\big \{} {\bf z}_{t} \log ( {\bf y}_{t} ) + (1-{\bf z}_{t}) \log (1-{\bf y}_{t}) {\big \}}\nonumber\\[-1pt]
&\hspace{40pt} + \alpha {\bf v}^{\mathsf T} {\bf L} {\bf v},
\end{align}
%\vspace{0pt}
%

\noindent where $\alpha$ is the regularization weight.
By approximating the frequencies of sound event occurrences ${\bf v}$ by $\sum_{t} {\bf y}_{t}$, we finally obtain the objective function as

%
%\vspace{0pt}
\begin{align}
E({\bi \Theta}) &= - \sum^{T}_{t=1} {\big \{} {\bf z}_{t} \log ( {\bf y}_{t} ) + (1-{\bf z}_{t}) \log (1-{\bf y}_{t}) {\big \}}\nonumber\\[-3pt]
&\hspace{40pt} + \alpha {\big (} \sum_{t=1}^{T} {\bf y}_{t} {\big )}^{\! \mathsf T} {\bf L} {\big (} \sum_{t=1}^{T} {\bf y}_{t} {\big )}\nonumber\\
&= - \hspace{-2.5pt} \sum^{M}_{m=1} \hspace{-1pt} \sum^{T}_{t=1} \hspace{-1pt} {\Big \{} \hspace{-1pt} z_{m,t} \log y_{m,t} \hspace{-1pt} + \hspace{-1pt} (1 \hspace{-1pt} - \hspace{-1pt} z_{m,t}) \log ( 1 \hspace{-1pt} - \hspace{-1pt} y_{m,t} ) \hspace{-1pt} {\Big \}}\nonumber\\[-3pt]
&\hspace{40pt} + \alpha {\big (} \sum_{t=1}^{T} {\bf y}_{t} {\big )}^{\! \mathsf T} {\bf L} {\big (} \sum_{t=1}^{T} {\bf y}_{t} {\big )}.
\end{align}
%\vspace{0pt}
%

\noindent Thus, we can detect sound events ${\bf y}_{t}$ while considering the co-occurrence of sound events.
Note that the proposed method can also be applied to any neural network system such as CNN, RNN, and CRNN-based systems when the output of a network is represented by ${\bf y}_{t}$.

\begin{table*}[t]
\small
\caption{Average performance of SED for each event in terms of F1 score}
%\vspace{0pt}
\label{tbl:result02}
\centering
\scalebox{0.818}[0.818]{
\begin{tabular}{lrrrrrrrrrrrr}
\wcline{1-13}
&&&&&&&&&&&&\\[-6pt]
\multirow{2}{*}{Event} & \multicolumn{1}{c}{(object)} & \multicolumn{1}{c}{(object)} & \multicolumn{1}{c}{(object)} & \multicolumn{1}{c}{(object)} & \multicolumn{1}{c}{(object)} & \multicolumn{1}{c}{bird} & \multicolumn{1}{c}{brakes} & \multicolumn{1}{c}{person} & \multicolumn{1}{c}{\multirow{2}{*}{car}} & \multicolumn{1}{c}{\multirow{2}{*}{children}} & \multicolumn{1}{c}{\multirow{2}{*}{cupboard}} & \multicolumn{1}{c}{\multirow{2}{*}{cutlery}}\\
& \multicolumn{1}{c}{banging} & \multicolumn{1}{c}{impact} & \multicolumn{1}{c}{rustling} & \multicolumn{1}{c}{snapping} &  \multicolumn{1}{c}{squeaking} & \multicolumn{1}{c}{singing} & \multicolumn{1}{c}{squeaking} & \multicolumn{1}{c}{breathing} &&&&\\
&&&&&&&&&&&&\\[-8pt]
\wcline{1-13}
&&&&&&&&&&&&\\[-9pt]
CNN & 0.00\% & 0.03\% & 0.02\% & 0.00\% & 0.00\% & 46.78\% & 3.58\% & 0.00\% & 59.32\% & 0.00\% & 0.00\% & 0.00\% \\
CNN-BiGRU & 0.00\% & 0.00\% & 6.45\% & 0.00\% & 0.00\% & {\bf 55.13\%} & 0.00\% & 0.00\% & 54.29\% & 0.00\% & 0.00\% & 0.00\% \\
CNN-BiGRU w/ GLR & 0.00\% & {\bf 0.83\%} & {\bf 16.81\%} & 0.00\% & 0.00\% & 39.54\% & {\bf 6.20}\% & 0.00\% & {\bf 60.63\%} & 0.00\% & 0.00\% & 0.00\% \\
\wcline{1-13}
\vspace{0pt}
\end{tabular}
}
\small
\label{tbl:each_event}
\centering
\scalebox{0.792}[0.792]{
\begin{tabular}{lrrrrrrrrrrrrr}
\wcline{1-14}
&&&&&&&&&&&&&\\[-6pt]
\multirow{2}{*}{Event} & \multicolumn{1}{c}{\multirow{2}{*}{dishes}} & \multicolumn{1}{c}{\multirow{2}{*}{drawer}} & \multicolumn{1}{c}{\multirow{2}{*}{fan}} & \multicolumn{1}{c}{glass} & \multicolumn{1}{c}{keyboard} & \multicolumn{1}{c}{large} & \multicolumn{1}{c}{mouse} & \multicolumn{1}{c}{mouse} & \multicolumn{1}{c}{people} & \multicolumn{1}{c}{people} & \multicolumn{1}{c}{washing} & \multicolumn{1}{c}{water tap} & \multicolumn{1}{c}{wind} \\
&&&& \multicolumn{1}{c}{jinging} & \multicolumn{1}{c}{typing} & \multicolumn{1}{c}{vehicle} & \multicolumn{1}{c}{clicking} & \multicolumn{1}{c}{wheeling} & \multicolumn{1}{c}{talking} & \multicolumn{1}{c}{walking} & \multicolumn{1}{c}{dishes} & \multicolumn{1}{c}{running} & \multicolumn{1}{c}{blowing}\\
&&&&&&&&&&&&&\\[-8pt]
\wcline{1-14}
&&&&&&&&&&&&&\\[-9pt]
CNN & 0.21\% & 0.00\% & 36.39\% & 0.61\% & {\bf 2.77\%} & 43.93\% & {\bf 20.41\%} & 0.00\% & 0.00\% & 0.07\% & 20.01\% & {\bf 54.95\%} & 0.06\% \\
CNN-BiGRU & 0.28\% & 0.00\% & 61.29\% & 0.00\% & 0.42\% & 43.22\% & 0.00\% & 0.00\% & 0.00\% & 11.39\% & 5.29\% & 33.91\% & 0.00\% \\
CNN-BiGRU w/ GLR & {\bf 14.16\%} & 0.00\% & {\bf 68.96\%} & {\bf 2.53\%} & 1.09\% & {\bf 49.66\%} & 0.00\% & 0.00\% & {\bf 0.01\%} & {\bf 48.88\%} & {\bf 33.82\%} & 41.62\% & {\bf 6.14\%} \\
\wcline{1-14}
\end{tabular}
}
%\vspace{3pt}
\end{table*}
%
%
%
%
%\vspace{0pt}
\begin{figure}[t]
%\vspace{0pt}
\centering
%\vspace{0pt}
\includegraphics[scale=0.49]{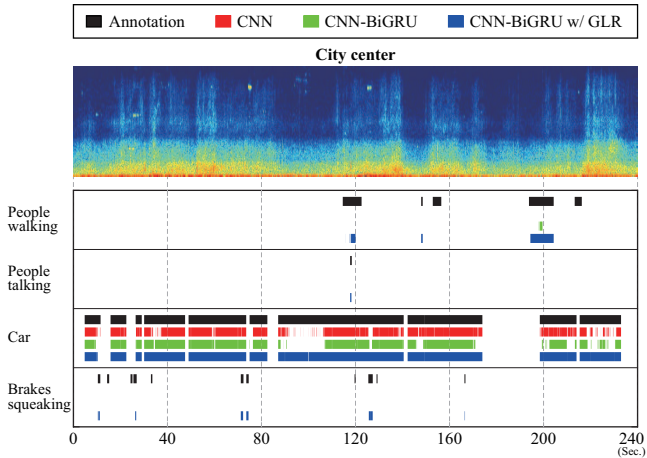}
%\vspace{0pt}
\caption{Annotations and event detection results for sounds recorded in city center. Only sound events occurring in the annotations are described.}
\label{fig:b098}
\end{figure}
%
%
%
%\vspace{0pt}
\begin{figure}[t]
%\vspace{0pt}
\centering
%\vspace{0pt}
\includegraphics[scale=0.49]{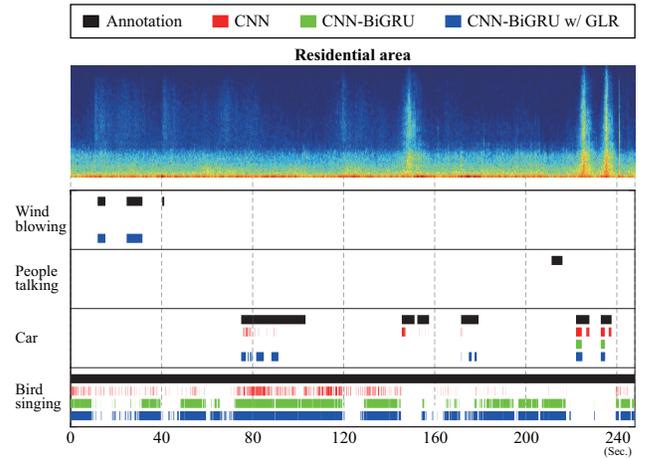}
%\vspace{0pt}
\caption{Annotations and event detection results for sounds recorded in residential area. Only sound events occurring in the annotations are described.}
\label{fig:a003}
\end{figure}
%
%
%
%------------------------------------------------------------------
%\vspace{0pt}
\section{Experiments}
\label{sec:experiments}
%\vspace{0pt}
%------------------------------------------------------------------
%- - - - - - - - - - - - - - - - - - - - - - - - - - - - - - - - - - - -
%\vspace{0pt}
\subsection{Experimental Conditions}
\label{subsec:conditions}
%\vspace{0pt}
%- - - - - - - - - - - - - - - - - - - - - - - - - - - - - - - - - - - -
We conducted evaluation experiments with conventional neural-network-based methods and the proposed method.
For the experiments, we constructed a dataset composed of parts of the TUT Sound Events 2016 and 2017 development datasets, and the TUT Acoustic Scenes 2016 development dataset \cite{Mesaros_EUSIPCO2016_01,Mesaros_DCASE2017_01}.
From the three datasets, we used sound clips including four acoustic scenes, ``home,'' ``residential area'' (TUT Sound Events 2016), ``city center'' (TUT Sound Events 2017), and ``office'' (TUT Acoustic Scenes 2016), with a total duration of 192 min of audio.
The experimental data include the 25 types of sound events listed in Fig.~\ref{fig:numofinstance}.
In this regard, because the original TUT Acoustic Scenes 2016 development datasets do not have sound event labels for the sound clips recorded in the office environment, we annotated them using the same procedure as that described in \cite{Mesaros_EUSIPCO2016_01} and \cite{Mesaros_DCASE2017_01}.
The experiments were conducted using the four fold cross-validation setup introduced in the TUT Acoustic Scenes 2016 and 2017 development datasets.

As the input of the system, the 64-dimensional log mel-band energy was used.
The adjacency matrix ${\bf A}$ was calculated by counting the number of co-occurring sound events in each sound clip over the training dataset and normalizing the result in the range from 0 to 1.
The adjacency matrix ${\bf A}$ used in the evaluation experiments is shown in Table~\ref{tbl:event_occurrence}.
After the model training, active sound events were predicted by thresholding the output ${\bf y}_{t}$ using an adaptive thresholding technique \cite{Xu_DCASE2017_01}.
The detection performance was evaluated in terms of the F1 score and error rate in the segment-based metrics \cite{Mesaros_AS2016_01}, in which the segment length is set to 40 ms.
The other recording conditions and experimental conditions are listed in Table~\ref{tab:Condition}, where the parameters of neural networks were selected by referring to \cite{Adavanne_DCASE2017_01}.
\begin{table*}[t!]
%\vspace{0pt}
\caption{Detection performance of sound events with respect to each acoustic scene}
\label{tab:result03}
%\vspace{4pt}
\small
\centering
\renewcommand{\arraystretch}{1.0}
\begin{tabular}{lrrrrrrrr}
\wcline{1-9}
\!&&&&&&\vspace{-9pt} \\
\multicolumn{1}{c}{\multirow{2}{*}{\bf Acoustic scene}} & \multicolumn{2}{c}{\bf City center} & \multicolumn{2}{c}{\bf Home} & \multicolumn{2}{c}{\bf Office} & \multicolumn{2}{c}{\bf Residential area}\\
\cline{2-9}\\[-9pt]
&F1 score & Error rate \ \ &\ \  F1 score & Error rate \ \ &\ \  F1 score & Error rate \ \ &\ \  F1 score & Error rate\\
\wcline{1-9}\\[-9pt]
CNN & 39.53\% & 0.255 & {\bf 16.12\%} & {\bf 0.112} & 34.61\% & 0.215 & 34.25\% & {\bf 0.227}\\
CNN-BiGRU & 38.84\% & 0.257 & 7.14\% & 0.149 & 63.43\% & 0.145 & {\bf 35.27\%} & {\bf 0.227}\\
CNN-BiGRU w/ GLR & {\bf 54.74}\% & {\bf 0.204} & 15.51\% & 0.151 & {\bf 74.87\%} & {\bf 0.109} & 28.44\% & 0.268\\
\wcline{1-9}
\end{tabular}
\vspace{-8pt}
\end{table*}
\begin{table*}[t!]
\centering
%\vspace{0pt}
\caption{Average SED performance for various training data sizes}
\label{tab:Result04}
%\vspace{0pt}
\small
\centering
\renewcommand{\arraystretch}{1.0}
\begin{tabular}{lrrrrrrrr}
\wcline{1-9}
&&&&&&&\vspace{-9pt} \\
Training data size\ &\multicolumn{1}{c}{Original}&\multicolumn{1}{c}{1/2}&\multicolumn{1}{c}{1/4}&\multicolumn{1}{c}{1/8}&\multicolumn{1}{c}{1/16}&\multicolumn{1}{c}{1/32}&\multicolumn{1}{c}{1/64}&\multicolumn{1}{c}{1/128}\\
\wcline{1-9}\\[-9pt]
CNN\ &34.17\%&30.71\%&34.68\%&31.10\%&33.49\%&27.03\%&27.35\%&9.28\%\\[1pt]
CNN-BiGRU\ &41.24\%&39.76\%&\textbf{40.79\%}&28.06\%&30.39\%&29.97\%&28.71\%&24.98\%\\[1pt]
CNN-BiGRU w/ GLR\ &\textbf{49.16\%}&\textbf{43.48\%}&39.93\%&\textbf{34.64\%}&\textbf{38.99\%}&\textbf{31.22\%}&\textbf{31.53\%}&\textbf{26.88\%}\\
\wcline{1-9}
\end{tabular}
\vspace{2pt}
\end{table*}
%
%
%- - - - - - - - - - - - - - - - - - - - - - - - - - - - - - - - - - - -
%\vspace{0pt}
\subsection{Overall Detection Performance of Sound Events}
\label{subsec:results_overall}
%\vspace{0pt}
%- - - - - - - - - - - - - - - - - - - - - - - - - - - - - - - - - - - -
Table~\ref{tab:Result01} shows the detection performances of CNN, CNN-BiGRU, and CNN-BiGRU with graph Laplacian regularization (GLR) in terms of the micro F1 score and error rate.
The results show that the proposed method considerably improves the SED performance in terms of both the F1 score and error rate.
In particular, the proposed method improves the average SED performance by 7.9 percentage points from that of the conventional CNN-BiGRU-based method in terms of the F1 score.

To investigate the detection results for sound events in more detail, we illustrate examples of annotations and the predicted results in Figs.~\ref{fig:b098} and \ref{fig:a003}.
The results also show that the proposed method detects sound events more accurately than the conventional methods.
In particular, the proposed method can detect co-occurring sound events with less overlook than the conventional methods.
For instance, the sound events ``car'' and ``brakes squeaking'' can be detected simultaneously by the method adopting graph Laplacian regularization, and the sound events ``wind blowing'' and ``bird singing'' can also be detected simultaneously.
On the other hand, the conventional methods cannot detect ``brakes squeaking'' and ``wind blowing'' events.
Thus, we conclude that graph Laplacian regularization based on the co-occurrence of sound events is a promising technique for SED.
%
%
%- - - - - - - - - - - - - - - - - - - - - - - - - - - - - - - - - - - -
%\vspace{0pt}
\subsection{Sound Event Detection Performance for Each Sound Event and Scene}
\label{subsec:results_each}
%\vspace{0pt}
%- - - - - - - - - - - - - - - - - - - - - - - - - - - - - - - - - - - -
We also examined the SED performance for each sound event.
The experimental results are shown in Table~\ref{tbl:result02}.
From these results and Table~\ref{tbl:event_occurrence}, it is also shown that the graph Laplacian regularization enables us to detect the co-occurring sound events more accurately, for example, the detection performance for the sound events ``car,'' ``brakes squeaking,'' and ``large vehicle'' improves compared with the conventional SED methods.
On the other hand, the detection performance for the sound events ``(object) banging,'' ``(object) squeaking,'' and ``cupboard'' does not improve.
This is because these sound events occur less frequently in the dataset, and the weights of edges corresponding to these sound events are small; thus, the graph Laplacian regularization had less effect on the detection of these sound events.

We then evaluated the detection performance for sound events for each acoustic scene.
Table~\ref{tab:result03} shows the average detection performance of sound events for each acoustic scene in terms of the macro F1 score.
In the acoustic scenes ``city center'' and ``office,'' the detection performance for sound events using the graph Laplacian regularization is higher than that of the conventional SED methods, whereas in the acoustic scenes ``home'' and ``residential area,'' the proposed method deteriorates the detection performance.
As shown in Fig.~\ref{tbl:event_occurrence}, many sound events occurring in the acoustic scenes ``home'' and ``residential area,'' such as  ``(object) banging,'' ``(object) squeaking,'' ``cupboard,'' and ``drawer,'' take small weights of edges.
Therefore, the adjacency matrix might not clearly represent the co-occurrence of sound events, leading to the degradation of event detection performance.
%
%
%
%
%- - - - - - - - - - - - - - - - - - - - - - - - - - - - - - - - - - - -
%\vspace{0pt}
\subsection{SED Performance Evaluation Using Various Training Data Sizes}
\label{subsec:results_dataset}
%\vspace{0pt}
%- - - - - - - - - - - - - - - - - - - - - - - - - - - - - - - - - - - -
We evaluated the impact of the amount of training data used in the proposed method on the SED performance.
In this experiment, we also used the same four fold cross-validation setup as in the other experiments without varying the amount of training data, for which we randomly selected from $2^{-1}$ to $2^{-7}$ of the data for model training in each fold.
The adjacency matrix ${\bf A}$ was calculated by counting the number of co-occurring sound events over the selected training data.
The other experimental conditions were also the same as those in the other experiments.

Table~\ref{tab:Result04} shows the average SED performance for various amounts of training data.
The results show that when a limited amount of training data is used, the difference in SED performance between the proposed and conventional methods becomes smaller; however, the proposed method still outperforms the conventional methods.
When a smaller amount of training data is used for constructing ${\bf A}$, the adjacency matrix ${\bf A}$ becomes sparse, and the regularization using Eq. (\ref{eq:term1}) is less effective for sound event modeling.
On the other hand, each edge on the graph contributes to each regularization term $A_{i,j} ( v_{i} - v_{j} )^{2}$ in Eq. (\ref{eq:term1}), and we can utilize part of the information on the sound event co-occurrence even if the adjacency matrix ${\bf A}$ is sparse.
Thus, we consider that the proposed method is still advantageous for SED.
%
%
%
%------------------------------------------------------------------
%\vspace{0pt}
\section{Conclusion}
\label{sec:conclusion}
%\vspace{0pt}
%------------------------------------------------------------------
In this paper, we proposed an SED method based on a neural network using graph Laplacian regularization with the co-occurrence of sound events.
Unlike conventional CNN or CNN-BiGRU-based SED methods, the proposed method can detect sound events with prior information on the co-occurrence of sound events.
The proposed method enables sound events to be modeled effectively and efficiently even if there are many types of sound events to model and relatively limited training data.
The experimental results obtained using the TUT Sound Events 2016 and 2017 datasets, and the TUT Acoustic Scenes 2016 dataset show that the proposed method improves the SED performance by 7.9 percentage points in terms of the segment-based F1 score.
The experimental results also show that the proposed method can detect sound events that tend to co-occur, such as the sound events ``car'' and ``brakes squeaking,'' more accurately than the conventional methods.
In the future, further studies are required to improve the SED performance for sound events that occur less frequently in the training dataset.
%
%------------------------------------------------------------------
%\vspace{0pt}
\section*{Acknowledgments}
%\vspace{0pt}
%------------------------------------------------------------------
This work was supported by JSPS KAKENHI Grant Number JP19K20304 and the NVIDIA GPU Grant Program.
%
% -------------------------------------------------------------------------
\small
\bibliographystyle{IEEEbib}
\bibliography{IEEEabrv,IEICE2019refs02,KeisukeImoto08}
% -------------------------------------------------------------------------
%
%
%
\end{document}